\documentclass[12pt,preprint]{aastex}

\newcommand{\Msun}{\rm M$_\odot$}
\newcommand{\rsun}{\rm R$_{\odot}$}

\shorttitle{Linking Planets to Dusty White Dwarfs} 
\shortauthors{Debes et al.}
\begin{document}
\title{The Link Between Planetary Systems, Dusty White Dwarfs, and Metal Polluted White Dwarfs}
\author{John H. Debes\altaffilmark{1}, Kevin J. Walsh\altaffilmark{2}, Christopher Stark\altaffilmark{3}}

\altaffiltext{1}{Space Telescope Science Institute, Baltimore, MD 21218}
\altaffiltext{2}{Southwest Research Institute, Boulder, CO }
\altaffiltext{3}{Department of Terrestrial Magnetism, Carnegie Institution of Washington, Washington, DC 20015}

\begin{abstract}
It has long been suspected that metal polluted white dwarfs (types DAZ, DBZ, and DZ) and white dwarfs with dusty disks possess planetary systems, but a specific physical mechanism by which planetesimals are perturbed close to a white dwarf has not yet been fully posited.  In this paper we demonstrate that mass loss from a central star during post main sequence evolution can sweep planetesimals into interior mean motion resonances with a single giant planet.  These planetesimals are slowly removed through chaotic excursions of eccentricity that in time create radial orbits capable of tidally disrupting the planetesimal.  Numerical {\em N}-body simulations of the Solar System show that a sufficient number of planetesimals are perturbed to explain white dwarfs with both dust and metal pollution, provided other white dwarfs have more massive relic asteroid belts.  Our scenario requires only one Jupiter-sized planet and a sufficient number of asteroids near its 2:1 interior mean motion resonance.  Finally, we show that once a planetesimal is perturbed into a tidal crossing orbit, it will become disrupted after the first pass of the white dwarf, where a highly eccentric stream of debris forms the main reservoir for dust producing collisions.  These simulations, in concert with observations of white dwarfs, place interesting limits on the frequency of planetary systems around main sequence stars, the frequency of planetesimal belts, and the probability that dust may obscure future terrestrial planet finding missions.

\end{abstract}

\keywords{planetary systems--white dwarfs--numerical techniques: N-body--asteroids}

\section{Introduction}
Hundreds of exoplanets have been discovered around a variety of stars, from small M dwarfs out to large G giant stars, to tight orbits around pulsars.  The conclusion is clear that planet formation is a robust process in our galaxy.  A corollary to this conclusion is the idea that observable evidence for planetary systems must therefore exist from early star formation to various stellar end states.  The challenge is to identify which phenomena are uniquely due to planets, and which are considered false positives.  When planets are discovered at each stage of stellar evolution, it will be additionally challenging to interpret the impact stellar evolution had on the surfaces, composition, and orbits of such planets.  Studying these physical processes tell us what the eventual fate of our own Solar System will be.

A very compelling case can be made for the presence of planetary systems around metal enriched white dwarfs and white dwarfs that possess strong infrared excesses.  Many studies with {\em Spitzer} and ground based optical spectroscopy have provided nearly twenty dusty white dwarfs with some sort of abundance estimate for the dusty material, either through detection of the 10\micron\ silicate feature, careful abundance analysis of narrow metal lines in the photosphere of the target white dwarf, or detection of gaseous emission lines \citep{zuckerman03,koester05,reach05,kilic06,gaensicke06,vonhippel07,farihi10,debes11}.  More recently, the Wide Field Infrared Survey Explorer (WISE)\citep{debes11b}, has shown that $\sim$1\% of WDs show dusty disks, confirming earlier work with Spitzer.  In contrast, nearly 20-25\% of all WDs show measurable metal accretion \citep[e.g.][]{zuckerman03}.

There have been two main conclusions from this excellent body of work: 1) that the material drizzling onto the surfaces of these white dwarfs is primarily similar to the inner terrestrial planets in our solar system, and 2) that planetesimals which are tidally disrupted best fit the observed location and composition of the dust disks \citep{zuckerman07,klein10,jura03}.

\citet{jura03} and \citet{jura07} have elegantly shown that most white dwarfs with infrared excesses require inner disk radii of just a dozen or so {\em white dwarf} radii, or about 0.1 R$_\odot$.  The extreme proximity to the white dwarf produces quite unusual conditions for a dust disk, and argues strongly for a tidally disrupted asteroid that has arrived from beyond a few AU, 
since anything at smaller distances during post main sequence evolution would be evaporated within the envelope of the evolving giant star.  Based on these arguments, it has generally been assumed that a planet several AU away from the WD is responsible for perturbing planetesimals into highly elliptical, tidally disrupting orbits.

A primary uncertainty in linking planetary systems to dusty and polluted white dwarfs is the exact dynamical mechanism that delivers planetesimals so close to the white dwarf.  \citet{alcock86}, in a prescient paper, tried to explain the presence of the metal polluted hydrogen WD (type DAZ) G74-7 with infrequent cometary impacts onto a white dwarf surface from Oort cloud analogues.  Further work was done looking at the dynamical stability of planetary systems during post-main sequence evolution as a possible driver for material into the inner system \citep{debes02}.  This scenario could reproduce the rough number of polluted white dwarfs assuming that as many as 50\% of planetary systems are unstable.  It also predicted that some planetary systems may be stable for up to 1~Gyr after a star turns into a WD, in accordance with the range of pollution observed for WDs.  Because dusty white dwarfs are observed at relatively late cooling ages when the WD has very low luminosity, processes that are initially efficient at creating dust during post main sequence evolution become negligible once a white dwarf is formed and is cooling \citep{dong10,bonsor10}.  Recently, \citet{bonsor11} investigated the frequency with which planetesimals in belts exterior to a giant planet might be perturbed into the inner system.  They found in their {\em N}-body simulations that a sufficient number of planetesimals can be perturbed to match the observed evolution of the dust accretion rate onto white dwarfs as a function of cooling age.  Two main assumptions however were used in that work: that the planetesimals must survive post main sequence evolution, and that roughly 10\% of perturbed planetesimals would be further deflected (presumably by a second interior planet) into tidally disrupting orbits.  Most planetesimals beyond the ice line (Kuiper belt type analogues) should be primarily icy.  They may significantly sublimate during post main sequence evolution before they can be perturbed.  Similarly, many interior planets that could further perturb outer planetesimals into white dwarf-crossing orbits may be engulfed during post main sequence evolution.  Both scenarios could be helped by eccentricity pumping of icy bodies further away from the inner system that don't see the mass loss of the central star as adiabatic \citep{veras11}.

While all dusty white dwarfs show evidence for metal accretion onto their surfaces, not all metal enriched white dwarfs show evidence of a dusty disk.  A correlation between accretion rate and the presence of dusty disks has been suggested \citep{vonhippel07,farihi09,zuckerman10}, with an accretion rate of $>$10$^{8.5}$ g/s consistent with the range above which dusty white dwarfs occur.  The delineation could be due to the average size of the perturbed planetesimal \citep{jura08}.  In this scenario, dusty disks only occur for larger than average disruptions, while smaller planetesimals that are disrupted are quickly turned to gas through mutual collisions and sputtering.   Evidence for the presence of multiple smaller disruptions causing gaseous disks has been shown with the discovery of circumstellar Ca gas absorption around WD 1124-293 \citep{debesub11}.  

In this paper we present a new scenario for perturbing planetesimals into highly eccentric orbits, thus strengthening the link between dusty white dwarfs and planetary systems.  We suggest in \S \ref{sec:mmp} that perturbations in eccentricity from interior mean motion resonances (IMMRs) with a giant planet roughly the mass of Jupiter are sufficient to create a steady stream of white dwarf crossing planetesimals.  In particular, we hypothesize that the 2:1 resonance is most efficient at driving white dwarf crossers, and that these asteroids are quickly tidally disrupted.  We use numerical simulations to follow the dynamics of the Solar System's asteroid belt  under the influence of post-main sequence evolution and perturbation by Jupiter in \S \ref{sec:numerical}.   Our model differs from the results of \citet{bonsor11} primarily in that we choose planetesimals that should be primarily rocky and survive post main sequence evolution \citep[see][]{jura08}, and we follow the dynamics of the planetesimals from perturbation to entering the tidal disruption radius of the white dwarf.  We also simulate the close approach of an asteroid to a white dwarf  in \S \ref{sec:numerical} to determine whether asteroids can be tidally disrupted quickly.  In \S \ref{sec:results} we confirm that a significant fraction of asteroids are perturbed into close encounters with a white dwarf, and that highly eccentric encounters between a small rubble pile asteroid and a white dwarf are sufficient to tidally disrupt the asteroid.  We compare the results of our models to the currently known population of dusty and polluted WDs in \S \ref{sec:comp} including estimating the asteroid belt masses necessary to produce the observed metal pollution in white dwarfs, and discuss our conclusions in \S \ref{sec:disc}.

\section{The Interior Mean Motion Perturbation Model}
\label{sec:mmp}
The Kirkwood gaps of the Solar System's asteroid belt are regions where asteroids are quickly removed due to gravitational interactions with planets.  Within IMMRs, an asteroid's eccentricity random walks until it undergoes a close encounter with a planet (or planets) and is ejected, collides with a planet, or collides with the central star \citep{morbi96,gladman97}.  This motion is limited to those bodies in an IMMR, whose width $\delta a_{max}$, can be roughly approximated by the maximum libration width for interior first order resonances in the restricted three body case and expanding the equations of motion for low eccentricities \citep{murraydermot}:

\begin{equation}
\label{eq:lib}
\delta a_{\rm max}=\pm\left(\frac{16}{3}\frac{|C_r|}{n}e\right)^{1/2}\left(1+\frac{1}{27j_2^2e^3}\frac{|C_r|}{n}\right)^{1/2}-\frac{2}{9j_2e}\frac{|C_r|}{n}a,
\end{equation}
where $n$ is the mean motion of the planetesimal and $e$ is the eccentricity of the planetesimal, $C_r$ is a constant from the resonant part of the disturbing function.  The constant $j_2$ comes from the resonant argument and determines which resonance is being used for a calculation,  and $\alpha$ is the ratio of the asteroid's semi-major axis to Jupiter's. The quantity $\frac{C_r}{n}$ is given by:
\begin{equation}
\frac{C_r}{n}=\mu\alpha |f_d(\alpha)|.
\end{equation}

\begin{figure}
\plotone{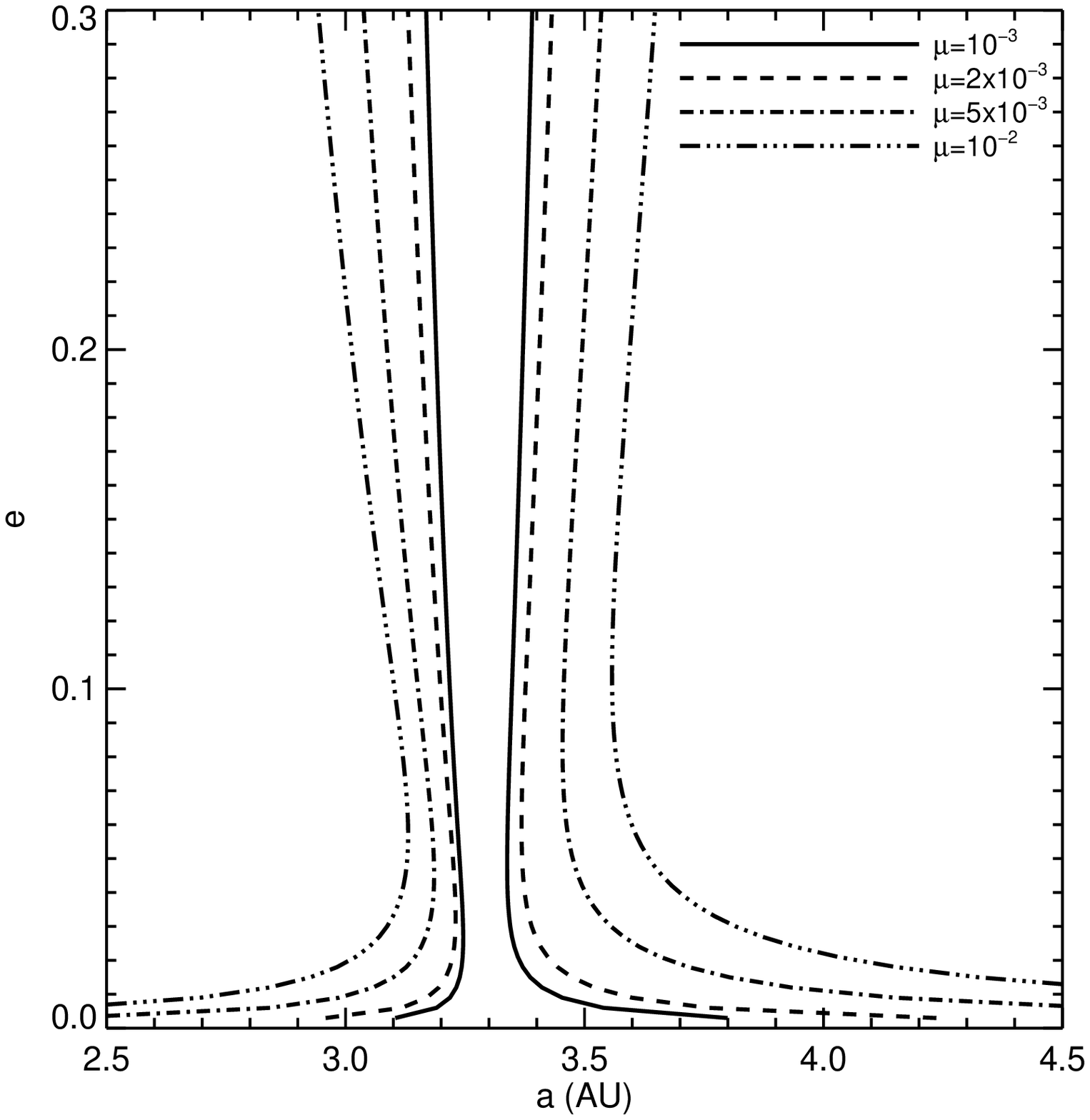}
\caption{\label{fig:f2} The evolution of the libration width about the 2:1 resonance as a function of mass lost from the central star (or as a function of planet mass ratio).  The libration width grows under increasing mass ratio, creating regions that can trap planetesimals that then later are perturbed out of the resonance.}
\end{figure}

For the 2:1 resonance, $j_2$=-1 and $\alpha f_d(\alpha)$=-0.749964 \citep{murraydermot}.  As can be seen in Equation \ref{eq:lib}, the maximum libration
width implicitly depends on the mass ratio $\mu$.  As a star evolves and loses mass, the mass ratio $\mu$ increases.  As a result, the width of the resonance $\delta a_{\rm max}$ increases and bodies previously exterior to the IMMR become trapped.  Just as in Hill stability and the stability 
of multi-planet systems \citep{gladman93,chambers96,debes02}, mass loss from the central star increases the perturbative influence of the planet.  Figure \ref{fig:f2} demonstrates the growth of $\delta a_{max}$ with a corresponding change in $\mu$ for the 2:1 resonance.

The 2:1 resonance in the Solar System appears to be the most efficient mechanism for scattering asteroids into Sun crossing orbits over Myr timescales.
  {\em N}-body simulations of asteroids injected into different resonances with Jupiter show a typical lifetime of a few million years (such as with the 3:1 resonance) to a few tens of millions of years for the 2:1 resonance \citep{gladman97}.  The 2:1 resonance is particularly attractive as a source for tidally disrupting asteroids around WDs because the timescale for asteroid removal is long (50\% removal at $>$100 Myr), and a significant percentage ($\sim$6.5\%) of objects in the 2:1 resonance are perturbed into Sun-crossing orbits. 

We now build on the tidal disruption scenario as envisioned by \citet{jura03}, which we call the Interior Mean Motion Perturbation (IMMP) model \citep{hoardbook}.  Planetary systems with one or more dominant giant planets possess planetesimal belts that are the remains of core accretion formation of terrestrial planets/giant planet embryos.  Over time these belts dynamically evolve through mean motion and secular resonances and lose mass to collisions, ejections, and star grazing orbits.  Many of the resonances are cleared out to $\sim \delta a_{\rm max}$.  During post main sequence evolution, smaller planetesimals migrate and evaporate in response to gas drag from the central star evolving off the main sequence, but a significant portion of large (r$>$1-3~km) planetesimals survive primarily unscathed \citep{jura08}.  Meanwhile, due to mass loss from the central star, the reservoir of planetesimals that are now vulnerable to eccentricity excursions from mean motion resonance perturbations has grown, either through trapping of medium sized planetesimals in resonance \citep{dong10,bonsor10} or through the growth of $\delta a_{\rm max}$.

Planetesimals that are on initially circular orbits interact gravitationally with a dominant giant planet and become perturbed into a highly elliptical orbit.  The perturbed planetesimals eventually obtain such high eccentricity that they are tidally disrupted by the central WD--the streams of debris from the initial disruption are primarily on similar orbits as their host planetesimal.  Over subsequent passes, more disruptions occur, spreading the material so that collisions increase.  Over many orbits these collisions damp down mutual eccentricity and form a coherent disk structure that evolves through a combination of further collisions and Poynting-Robertson drag until the dust becomes optically thick.  At this point the disk evolves viscously until it becomes optically thin, or all dust is sputtered into gas from grain-grain collisions and the debris from other asteroids \citep{jura08}.

\section{Numerical Methods}
\label{sec:numerical}
In order to test the above IMMP model, we will demonstrate that planetesimals 1) entered a highly eccentric orbit within the tidal disruption radius, and 2) disrupted within the expected disruption radius.  Ideally our simulations would also follow the disruption at later times to see if a dust disk forms that matches observations of known dusty white dwarfs, but this is computationally intensive with the techniques we use in this study and is saved for future work.

To demonstrate that the IMMP model satisfies the first criterion, we dynamically simulated the response of the Solar System's asteroid belt to post main sequence evolution.  We chose the Solar System primarily because it should be a decent proxy for a planetary system where only one planet dominates the inner planetesimal region, and should not be subject to any biases associated with an incorrect treatment of Gyr of dynamical evolution between a planet and its planetesimal belt.
Our simulations follow the largest asteroids with well known radii--the lower limit to our population is larger than the minimum post main sequence survival radius ($\approx$10~km) for distances of a few AU.  We do not follow smaller asteroids that may be trapped in IMMR resonances during the late stages of post main sequence evolution.  If we assume that WDs have similar mass asteroid belts, our simulations should represent a lower limit to the amount of material tidally disrupted and accreted onto WDs.   

We also simulated the tidal disruption of a small asteroid by a white dwarf using the {\tt pkdgrav} code.  We modified the code to treat the approach of a small asteroid to within $<$1\rsun\ at various close approaches to determine whether criterion 2) was satisfied, drawing on the orbital elements inferred from our {\em N}-body simulations.

\subsection{The Solar System as a Laboratory for Dusty White Dwarfs}

In order to follow the behavior of the 2:1 resonance, we performed numerical simulations of large Solar System asteroids in which mass loss from the central star is included.  We have taken ten MERCURY simulations using a Burlirsch-Stoer integrator with an adaptive timestep \citep{chambers99} of 710 Solar System asteroids including Jupiter to determine if any follow the IMMP model.  

The asteroids were chosen to have radii R$>$50~km at all orbital semi-major axes and R$>10$~km with perihelia $>$3~AU based on the latest known asteroid data compiled by E. Bowell\footnote{ftp://ftp.lowell.edu/pub/elgb/astorb.html} and the assumption that the Sun will sublimate asteroids smaller and closer than this during post main sequence evolution \citep{schroeder08,jura08}.  The mass of the Sun was slowly removed over 1000~yr using the equation

\begin{equation}
M_\odot(t) = 1.0-0.46 \left[\left(\frac{t}{t_{\rm stop}}\right)^2-2\left(\frac{t}{t_{\rm stop}}\right)^3\right],
\end{equation}

reaching a mass of 0.54 \Msun\ \citep{schroeder08}.  Asteroids were removed if they strayed within 1 solar radius and considered tidally disrupted.  

Ten simulations were run for 100~Myr.  Four further simulations were performed for 200 Myr and three separate simulations using a Bulirsch-Stoer integrator from the hybrid symplectic integration package described in \citet{stark08} with a nominal step size of 1/40th of an orbit of Jupiter and adaptive timesteps were run for 1~Gyr to test whether the perturbation declines over timescales comparable to the cooling time of most white dwarfs.   The 1~Gyr simulations were also used as an independent test to confirm the general behavior seen with the MERCURY code.  The timescales we cover correspond to a large fraction of the cooling age for observed dusty and polluted white dwarfs.  For example, a $\log{g}$=8 white dwarf has an effective temperature of $\sim$18500~K at a cooling age of 100~Myr, $\sim$15000~K at 200~Myr, and $\sim$8200~K at 1~Gyr.  For comparison, all but one of the known dusty WDs have T$_{eff}$ $>$8200~K and 67\% of {\em Spitzer} observed DAZs have  T$_{eff}$ $>$8200~K.

\subsection{Modeling the Tidal Disruption of a Small Planetesimal}
The tidal disruption is simulated using a "rubble pile" model, based on a collection of 5000 hard, spherical particles. The simulations use the {\em N}-body code {\tt pkdgrav}, originally used for cosmological simulations, to provide a very fast, parallelized tree for computing inter-particle gravity \citep{richardson00}. The rubble pile simulations also take advantage of its collision resolution, in which particle collisions are detected and resolved according to a number of adjustable physical parameters, namely the coefficients of restitution (both tangential and normal). The body is then held together by its own self-gravity with particle collisions preventing a collapse. Previously this code has been used in numerous "rubble pile" asteroid models, including collisions \citep{leinhardt00,leinhardt02}, tidal disruptions \citep{richardson98,walsh06}, and also for planet formation simulations \citep{leinhardt05,leinhardt09}. Tidal disruption simulations have proven insensitive to numerous physical parameters, such as the tangential and normal coefficients of restitution, as the rotation imparted by the encounter and the "depth" of the encounter with the planet dominate the outcome \citep{richardson98,walsh06}.

The modeled tidal disruption consists of a 5000 particle body, having a single close encounter with a point mass of 0.5 \Msun. Such a point mass has a Roche radius of 89 R$_{\rm WD}$, or about 0.9 R$_\odot$, 695,000 km.  We chose orbits with semi-major axes of 4.77 AU, based on the orbital elements from one of the tidally disrupted asteroids in our N-body simulations.  The semi-major axis is larger than one would expect for a Solar System asteroid primarily due to the fact that asteroid orbits will expand in response to the mass loss from the central star.  The asteroid was modeled with different eccentricities to produce encounters inside the Roche limit at 80, 75, 70, 65, and 60 R$_{\rm WD}$. The progenitor was not spinning on the initiation of the encounter, and was only simulated for a single encounter. The simulation was well resolved, with a duration of 25,000 timesteps with each timestep representing 1e-5 yr/2$\pi$, or about 50 sec. At the end of each simulation the state of the aggregate was analyzed, and individual orbital elements for each ``clump'' were produced, along with the statistics for the number and size distribution of fragments.

\section{Results}
\label{sec:results}
\subsection{The Solar System}
\begin{figure}
\plotone{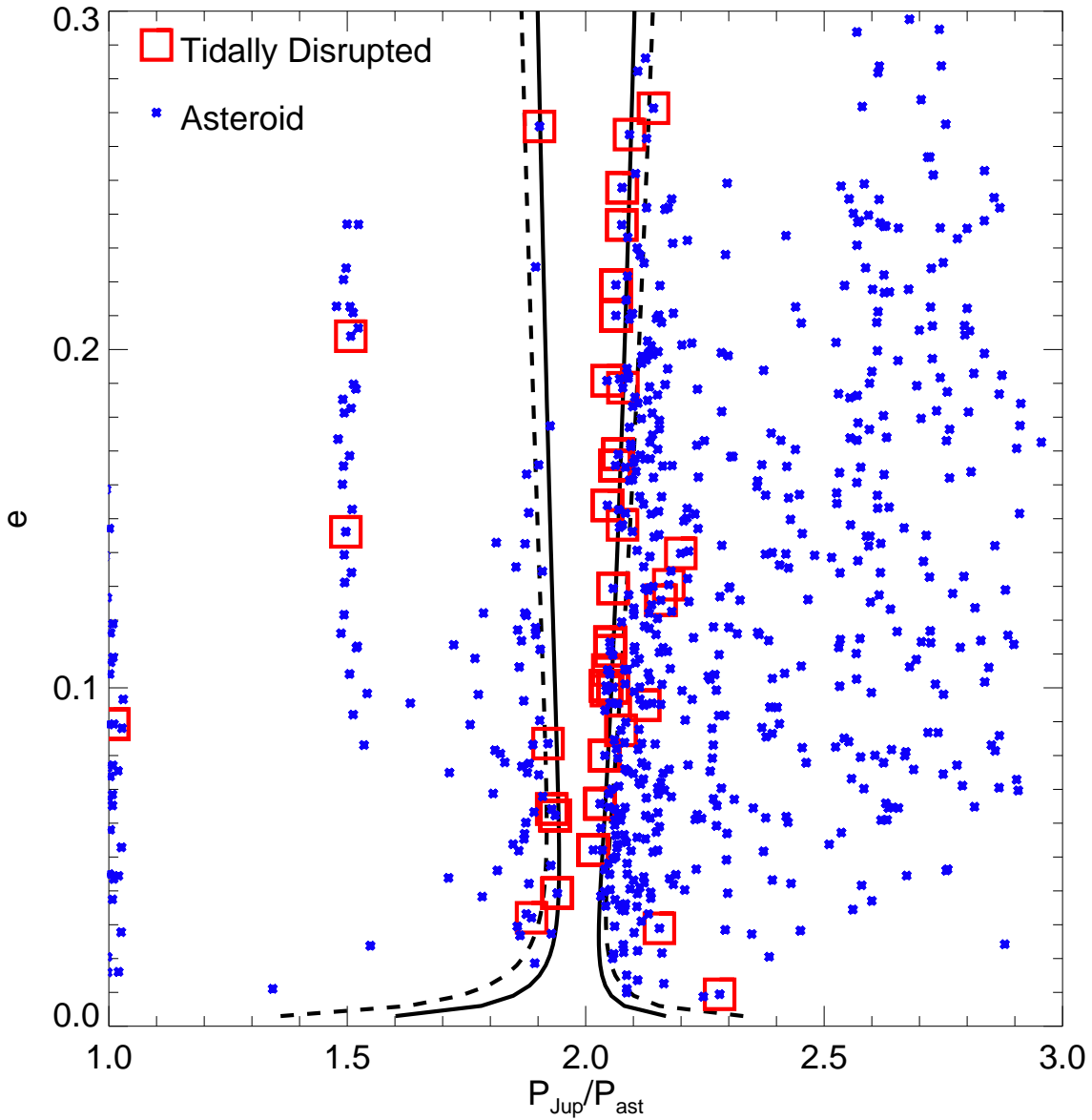}
\caption{\label{fig:f3} Summary figure of our {\em N}-body simulations of the Solar System asteroid belt under the influence of post-main sequence evolution.  Asteroids are plotted as a function of $e$ vs. $a$ for their initial orbital elements.  Overplotted are the libration widths for the 2:1 resonance before post-main sequence evolution (solid line) and after the Sun loses half of its mass (dashed line).  Small asterisks are the asteroids used in our simulations and the squares are asteroids that tidally disrupted in at least one of our 17 simulations.  The majority of the tidal disruptions occur around the edge of the new libration width to the 2:1 resonance after the central star's evolution.}
\end{figure}

Out of all the IMMP simulations, roughly 2\% of the modeled asteroids strayed within the tidal disruption radius, with the majority disrupting within the first few hundred Myr.  The resulting asteroids that were tidally disrupted are plotted as a function of their initial semi-major axis $a$ and eccentricity $e$ in Figure \ref{fig:f3}.  Overplotted are calculations of the libration width before and after post-main sequence evolution.

\begin{figure}
\plotone{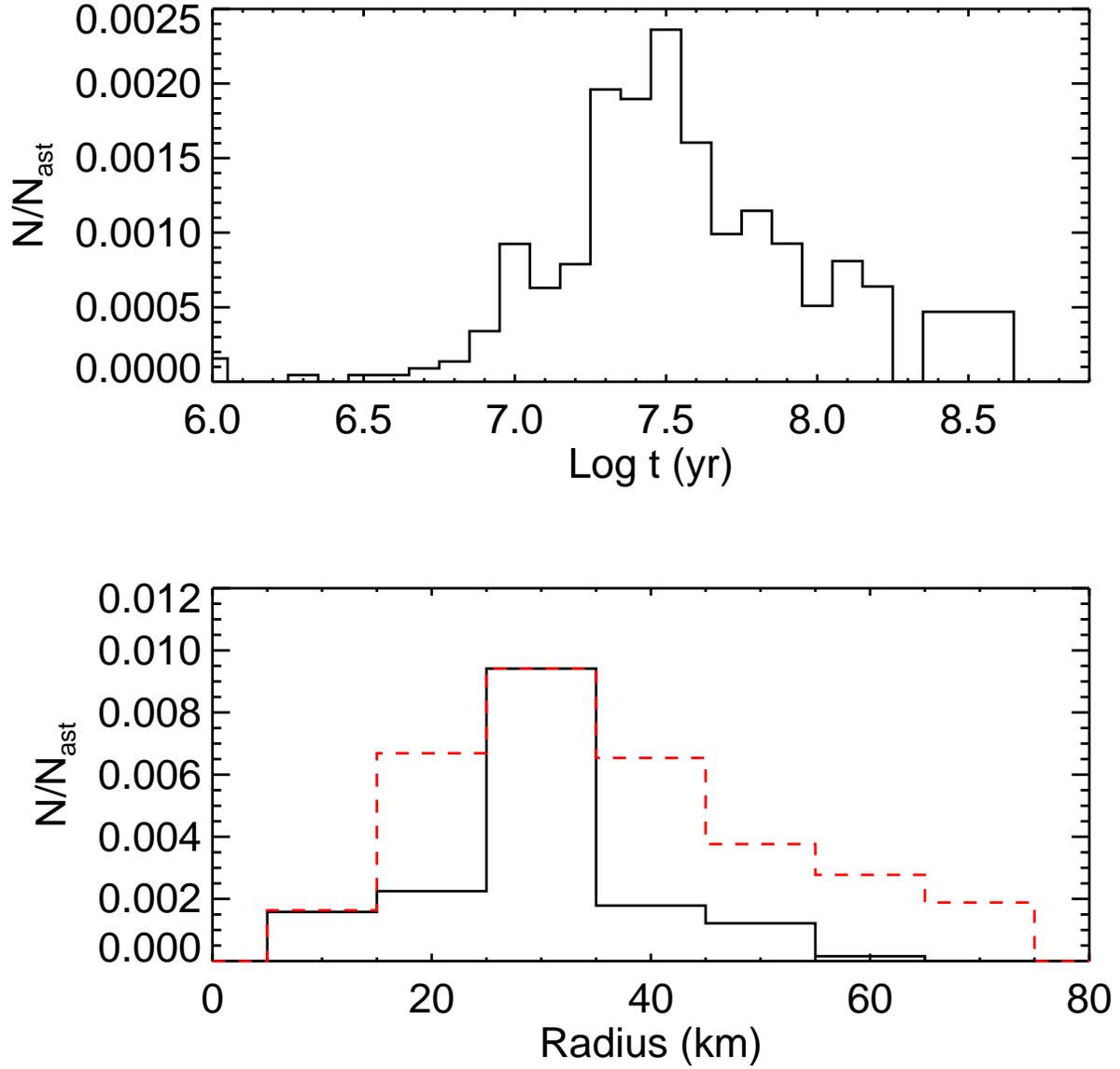}
\caption{\label{fig:f4} (top) The number of asteroids tidally disrupted as a function of time, normalized to the initial number of asteroids in the belt. (bottom) The size distribution of asteroids tidally disrupted in our simulations, normalized to the initial number of asteroids in the belt.  The red dashed line shows the initial distribution of asteroid radii normalized by the total number of asteroids used in our simulations, scaled by a factor of 0.0352.}
\end{figure}

In the 100~Myr simulations, one asteroid impacted approximately every 10~Myr.  From 100 to 200~Myr, the average time between impacts increased to 71~Myr, implying a near power law drop in the  impact frequency on the white dwarf as time went on.  In the 1~Gyr simulations, only one asteroid per simulation was disrupted beyond 200~Myr.  In order to compare across simulations, the number of asteroids disrupted as a function of time were normalized by the total number of asteroids simulated to the per asteroid frequency and binned logarithmically in time.  Figure \ref{fig:f4} shows the normalized number of asteroids disrupted as a function of time for all the simulations.  Taken as a whole, this curve would represent the average probability at any given time of an asteroid becoming tidally disrupted for the size distribution we use.   There is a peak at $\sim$30~Myr, which is consistent with a white dwarf T$_{eff}\sim$24000~K.  The hottest dusty white dwarfs are slightly cooler than this value and also represent some of the largest accretors of material, implying dust rich disks close to where the peak of perturbations occur in our simulations.  Beyond this peak, the frequency drops roughly as $t^{-0.6}$, but this is uncertain due to the small number statistics of our simulations.

We can compare these results to \citet{malhotra10}, who calculated the diffusion of solar system asteroids (with D$>$30~km) out of the asteroid belt due to perturbations from the giant planets.  They used second-order mixed variable symplectic mapping \citep{wisdom91,saha92} to integrate test particles between Mars and Jupiter to within 1~AU and then MERCURY with its hybrid integrator to perform their integrations within 1~AU.  For \citet{malhotra10}, 18\% of those asteroids perturbed into the inner Solar System impacted the Sun.  For our IMMP simulations, 17\% of the asteroids that were perturbed out of the asteroid belt also strayed to within 1~\rsun, our adopted tidal disruption radius.  \citet{malhotra10} also found that between 1-4~Gyr, roughly 0.4\% of asteroids in their simulations impacted the Sun, suggesting that even at later times, a significant flux of asteroids can be tidally disrupted.  Similarly, the late time dynamical evolution of the asteroid belt could be approximated by a power-law decline in the total number of asteroids close to $t^{-1}$, similar to the decline we see in our own impact frequency, suggesting that the rate of impacts decline with the total mass of the asteroid belt \citep{jura08}.

As noted previously in \citet{bonsor11}, one can determine an accretion rate $\dot{M}$ onto WDs based on the fraction of asteroidal belt mass scattered ($f_{\rm SI}$), disrupted ($f_{\rm TD}$), and eventually accreted ($f_{\rm acc}$), as well as the mean time of disruption $<t_{\rm TD}>$:

\begin{equation}
\label{eq:bonacc}
\dot{M}_{\rm metal} = \frac{f_{\rm acc} f_{\rm TD} f_{\rm SI} M_{\rm belt}}{<t_{TD}>}.
\end{equation}

In reality, the accretion rate onto the white dwarf surface is determined by the accretion of the dusty or gaseous disks that are generated from the asteroid tidal disruption.  Determining the precise evolution of a WD dust/gas disk is beyond the scope of this paper, so we assume an average accretion rate from the disk is given by $<t_{TD}>$.  We can determine accretion rate vs. time for our simulations if we can calculate a mass disrupted.  We took the asteroids tidally disrupted in our simulations normalized by the total number of asteroids sampled and calculated their mass using their measured radii, assuming a mean density of 4~g cm$^{-3}$.  This gives us the total mass of our asteroids normalized by the mass of the Solar System's asteroid belt ($M_{\rm belt}=3.6\times10^{24}$~g) \citep{krasinsky02}.  At each disruption time we determined the average mass mass accretion onto the white dwarf by setting $<t_{TD}>$ equal to the average time between the previous and next disruption, or the end time of the simulation and assuming that $f_{\rm acc}$ was of order unity.  This mostly follows the same procedure as \citet{bonsor11}, with the exception that we implicitly determine the quantity $f_{\rm TD} f_{SI} M_{\rm belt}$ by following individual disruptions, and we assume that $f_{\rm acc}$ is more efficient. 

\begin{figure}
\plotone{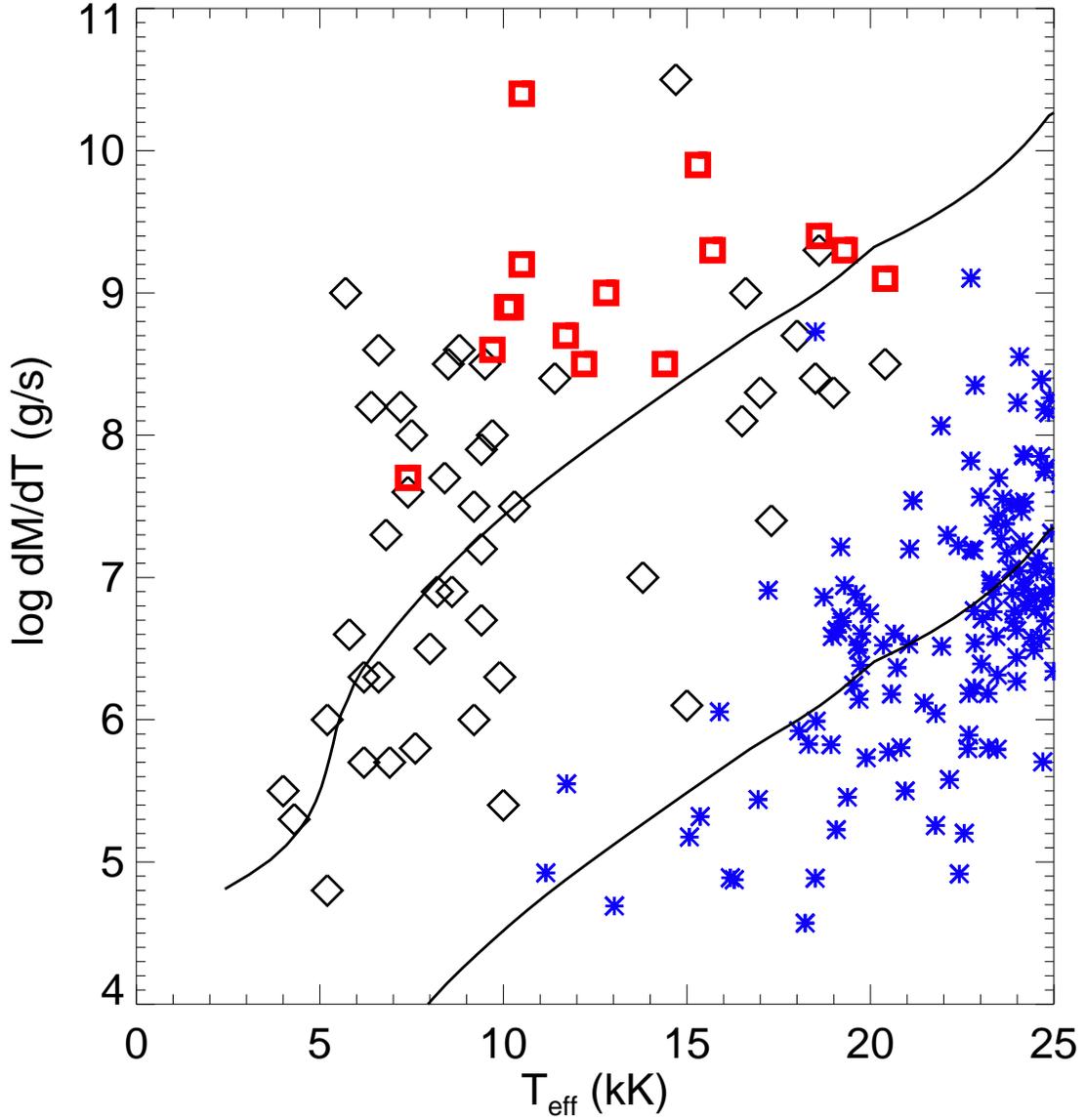}
\caption{\label{fig:f5} Observed $\dot{M}$ vs. T$_{eff}$ for both dusty (squares) and polluted (diamonds) WDs.  Asterisks represent the inferred instantaneous accretion rate from all of our simulations using Equation \ref{eq:bonacc}, while the solid curves represent a power-law fit to the simulated accretion rates as a function of time with a Solar System asteroid belt (lower curve) and an asteroid belt with 820 times the mass of the Solar System asteroid belt (upper curve).}
\end{figure}

 In principle the results of our simulations can be scaled to any starting asteroid belt mass by the total mass of asteroids, assuming a similar dynamical architecture of a dominant giant planet with an interior belt of asteroids following a similar size distribution to that of the Solar System's asteroid belt.  Figure \ref{fig:f5} shows our results for the accretion evolution of the Solar System compared to observed dusty and metal polluted white dwarfs as collected in \citet{farihi09,farihi10}.  For the figure, we have converted the time covered by our simulations into corresponding $T_{eff}$ assuming a hydrogen WD with $\log$~g=8.0.  Looking at all disruptions in our simulations, we take a power-law fit to our results and find that the mass accretion rate drops as t$^{-2}$.  We can extrapolate this relationship between time and accretion rate to the oldest WDs.  This is shown as the solid line in Figure \ref{fig:f5}.  The inferred accretion rates depend on our assumptions in the following way:

\begin{equation}
\label{eq:assump}
\dot{M}_{\rm metal} = \dot{M}_{\rm metal,o}\left(\frac{f_{\rm acc}}{1}\right)\left(\frac{M_{\rm belt}}{3.6\times10^{24}~g}\right).
\end{equation}

Our results suggest that the required mass in asteroids around other white dwarfs might have been much larger than our own Solar System, by as much as a factor of 10$^{3}$.  However, several uncertainties exist in our model.  The biggest uncertainty for this model is the exact location of the giant planet and how the timescale for asteroid perturbation out of the 2:1 resonance might scale with planetary architecture.  We also haven't accounted for how the accretion rate might evolve over the lifetime of a gaseous or dusty disk caused by the tidal disruption of an asteroid.  One would expect orders of magnitude higher accretion rates soon after a tidal disruption, with a power-law or exponential decay in accretion rate as mass is lost from the disk of material that forms from a tidal disruption--this effect would be balanced by the fraction of a WD's cooling age over which this high accretion phase might occur.  Such modeling would require a more careful analysis of how such disks evolve and the main mechanism for accretion onto the WD.  Recently, it has been shown that Poynting-Robertson drag may drive accretion in some cases, but cannot account for all of the accretion observed \citep{rafikov11a,rafikov11b,xu}.  

It is also unclear how tidally disrupted streams of material will evolve into dusty/gaseous disks and over what timescales this process occurs.  If the relative velocities of collisions are too high close to the WD, dust grains will suffer evaporative collisions and might not be efficient at leaving material close enough to accrete onto the WD.  Similarly, if relative velocities are too high, the tidally disrupted streams of material may not be able to collisionally damp into a disk with low enough eccentricity to efficiently accrete \citep{shannon11}.  Our sample of asteroids encompasses most of the largest asteroids present in our Solar System, but is missing most asteroids with radii $<$20~km.  Therefore it is possible that our implied tidal disruption rates could be higher if we take these objects into account, or other planetesimal belts have different size distributions such as what might be expected if a large number of planetesimals are trapped in resonances during post-main sequence evolution \citep{dong10}.  These uncertainties affect many of our assumed model values for the efficiency with which tidally disrupted material is accreted onto the WD and the average timescale between tidal disruptions.  Further work on these uncertainties through better observations of dusty disks, the modeling of WD disks, and longer simulations of dynamical perturbations with more particles could help to better predict how efficient the IMMP model is at delivering material close to a WD.  

\subsection{Tidal Disruption}
\label{sec:tidal}

In all of our tidal disruption simulations, the asteroid was significantly affected by the tidal forces of the central white dwarf, with varying degrees of disruption after the first pass.  As the periastron decreased, more violent disruptions occurred, with more mass being generated in smaller bodies.

\begin{figure}
\plotone{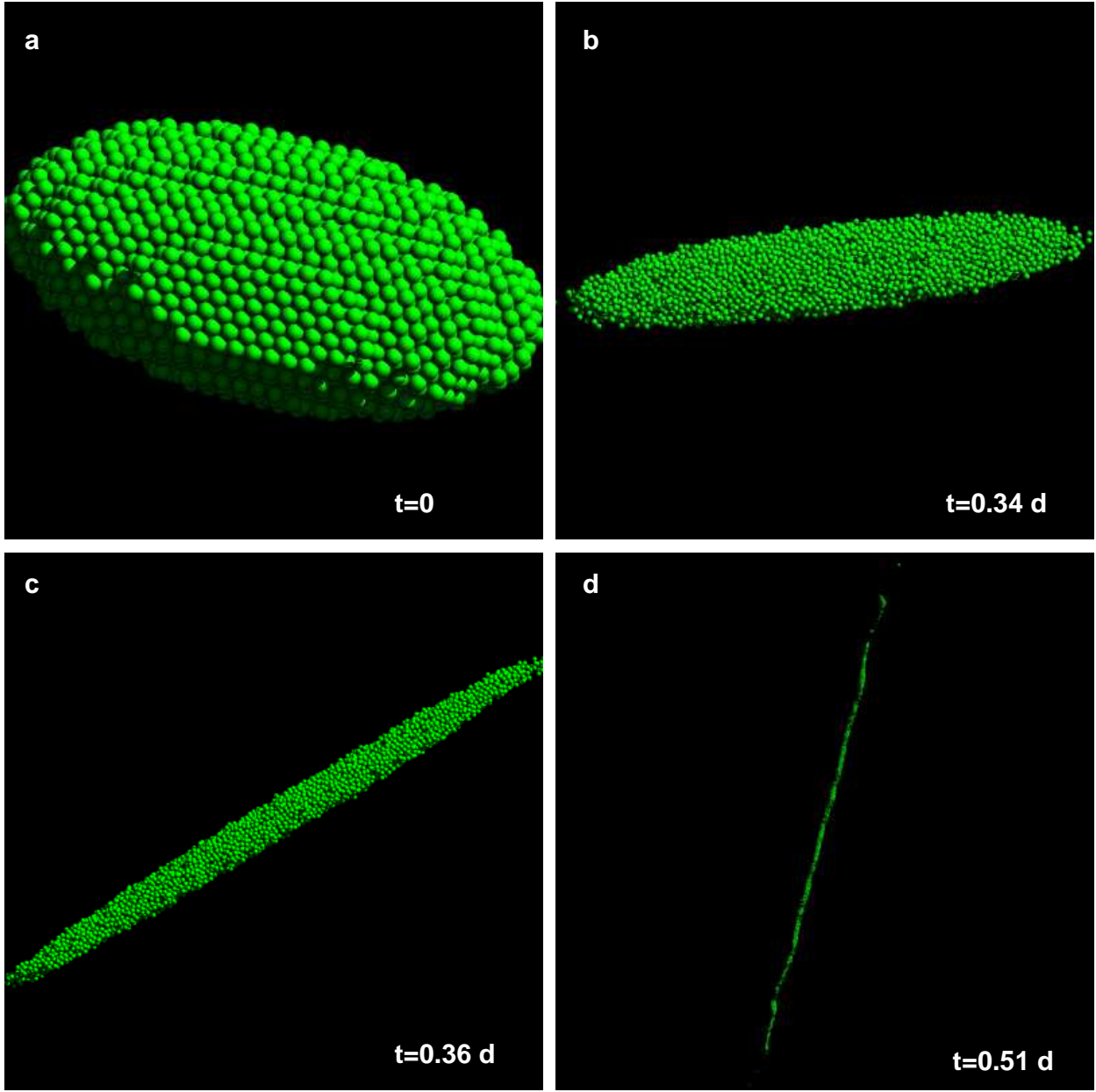}
\caption{\label{fig:f7} Snapshots of rubble pile simulations of an asteroid being tidally disrupted by a white dwarf.}
\end{figure}

Figure \ref{fig:f7} shows a snapshot of one asteroid disruption for a close encounter to 60~$R_{\rm WD}$.  The asteroid is elongated at the early stages of the disruption and quickly spreads out over 2~AU from head to tail of the train of debris after the first pass.  Despite this, the disrupted stream is rather similar in its orbital elements.
This tight stream is reflected for all the encounters and the orbital elements of the fragments were tightly centered around the original orbit of the incoming asteroidal body.  Further work needs to be done to determine whether these fragments will settle through mutual collisions and further tidal disruptions into a more circularized disk.  This suggests a phase of evolution, perhaps lasting many orbital timescales for the fragments, where dust is in an elliptical distribution.  In fact, observations of dusty white dwarfs with emission lines from gaseous disks show evidence of ellipticity \citep{gaensicke08}, though not at the level implied by our simulations, where the eccentricity of the debris stream is in excess of 0.99.  Over many orbital timescales, the fragments will spread and precess at slightly different rates and start colliding.  The exact evolution of the system will depend on the timescale for fragments to be directly perturbed onto the WD surface by the giant planet and the timescale for mutual collisions and a dynamical cooling of the systems.  Any dynamical cooling due to collisions could pull fragments out of resonance and allow them to survive longer to collide with other fragments, generating more dust.  The timescales and exact evolution of highly eccentric disks, like what we would expect from a tidal disruption, is an open question.

\begin{figure}
\plotone{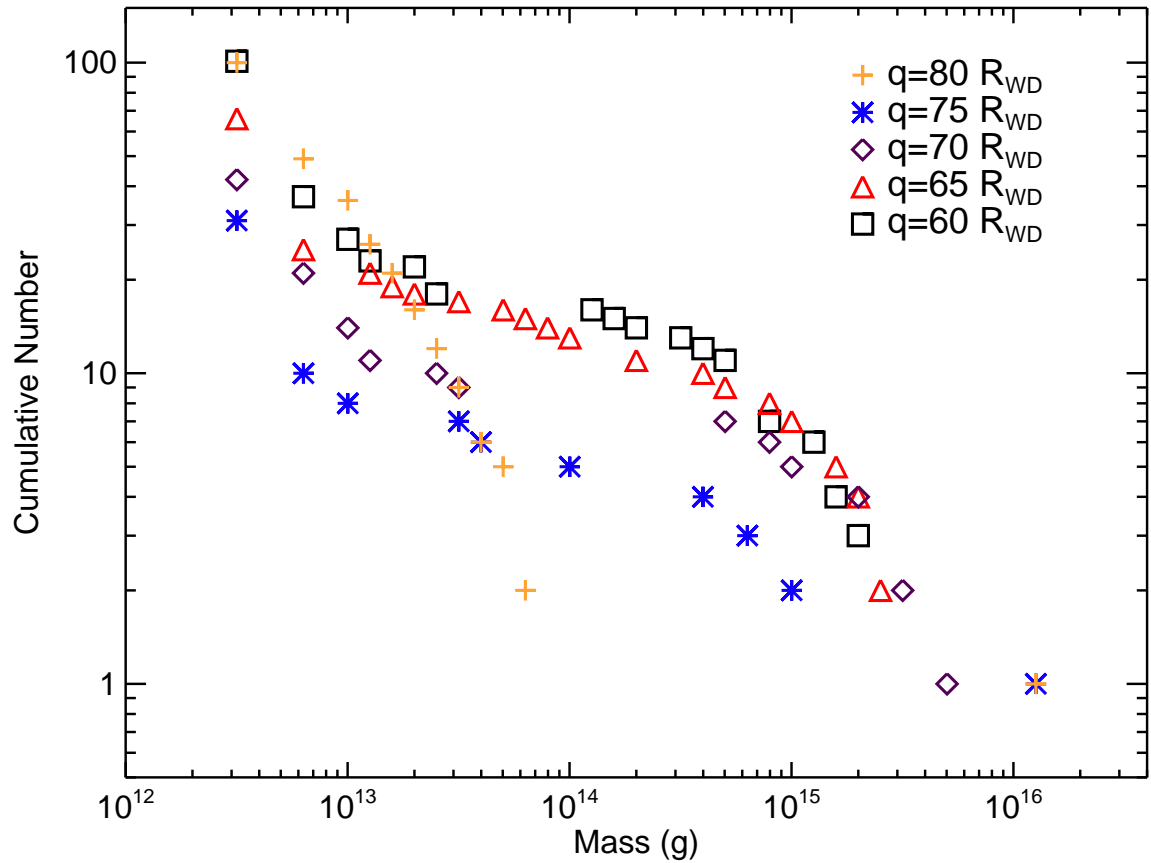}
\caption{\label{fig:f8} Cumulative size distributions of asteroid fragments after initial tidal disruption.  As the asteroid approaches closer to the WD, more fragments are generated.   The distance of close approach in WD radii is marked for each simulation.}
\end{figure}

Figure \ref{fig:f8} shows the cumulative number distribution of fragments as a function of mass.  From a close approach distance of 60 to 80 R$_{\rm WD}$.  Even after a single pass, several fragments were generated, with an increasing number of smaller fragments as the radius of close approach occurred.  In all of our simulations, there was a negligible amount of mass lost, suggesting that the initial disruption of the asteroid will not show in the WD photosphere until dust and material accreted through viscous evolution.

\section{Comparison of the IMMP model to the observed WD population}
\label{sec:comp}

We can compare the results of our IMMP model to observed metal polluted and dusty white dwarfs.  This is possible through the extensive compilation of {\em Spitzer} observed metal enriched white dwarfs presented in \citet{farihi09} and \citet{farihi10}.  We took these populations and divided them to investigate how each were distributed by WD cooling time and the total age of the system.

\begin{figure}
\plotone{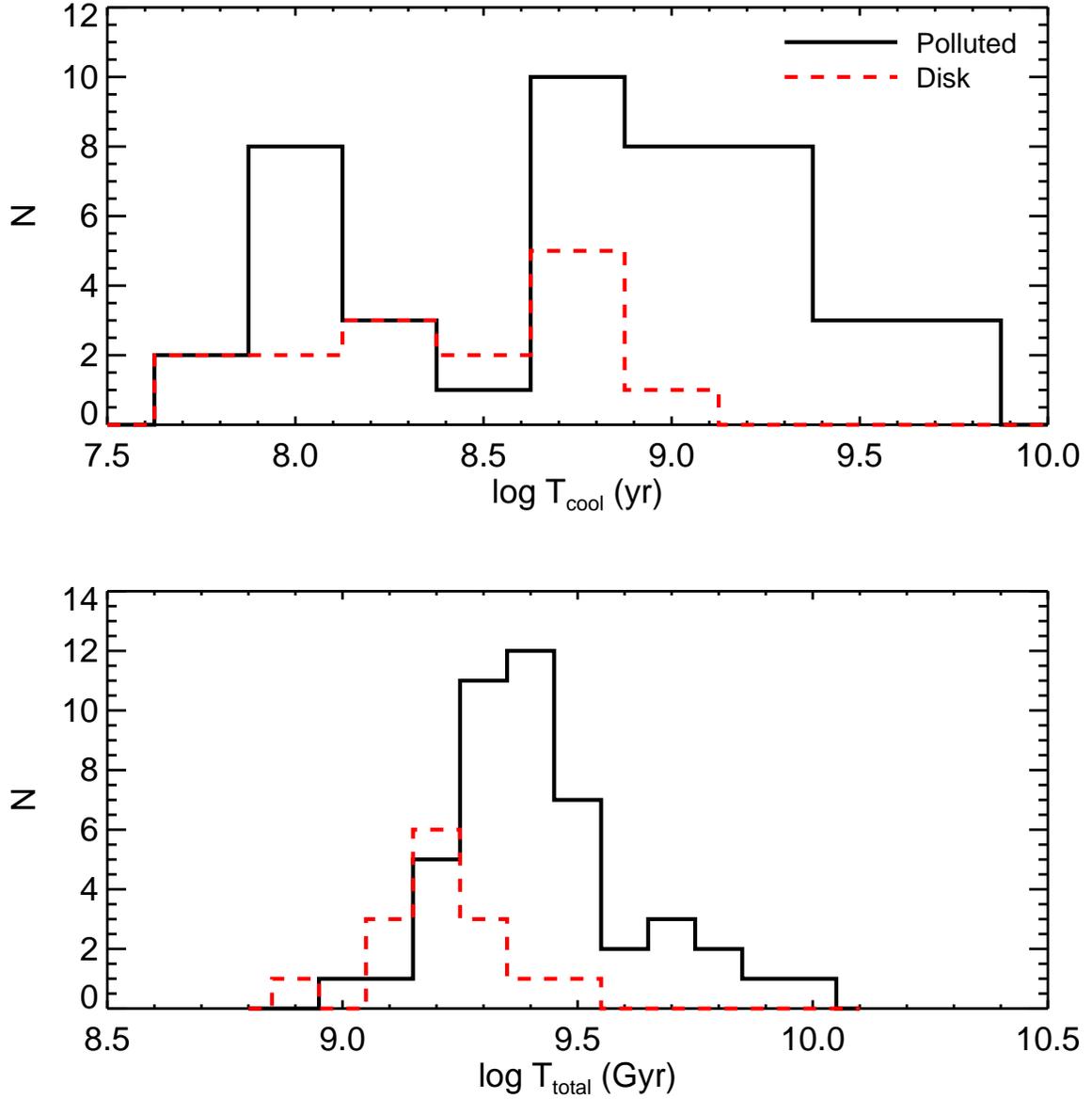}
\caption{\label{fig:f1} (top) Histograms of metal-rich WDs with no disks (solid line) and with disks (dashed line) as a function of cooling age ($t_{cool}$).  (bottom) Histograms of the same types of WDs but as a function of total age, $t_{cool}+t_{MS}$.} 
\end{figure}

The top panel of Figure \ref{fig:f1} shows the cooling age for the disk/non-disk populations.  There appears to be a bimodal population for non-disk systems that cluster around 10$^8$~yr and 10$^{9}$~yr, hinting at possibly two different mechanisms, or at least one mechanism that has two characteristic timescales after post-main sequence evolution.  If we interpret this in light of the IMMP model, these two timescales could represent the peak we observed at 30~Myr, but scaled to a longer characteristic timescale for perturbation, such as for a more widely separated planet.  The bimodality could also be a selection effect, since a smaller number of hotter WDs have been observed for metal accretion and small number statistics dominate.   There is not such a clear bimodality for the disk systems.  The median cooling age of the non-disk systems is 0.9~Gyr while the median cooling age for disk systems is 0.4~Gyr.

We also calculate the total age of each white dwarf, $t_{\rm tot}$=$t_{\rm MS}$+$t_{\rm cool}$.  The quantity $t_{\rm MS}$ can be determined first by inferring an initial mass for each WD using an empirical initial-final mass function \citep{williams09}:

\begin{equation}
\label{eq:initial}
M_{\rm final}=\frac{\left(M_{\rm initial}-0.339\right)}{0.129}
\end{equation}

The main sequence lifetime is then given by $t_{\rm MS}=10M_{\rm initial}^{-2.5}$.  
The bimodality of the non-disk systems now disappears, instead the distribution shows a peak around 2.3~Gyr, with a high age tail.  The disk systems are on average younger, with a peak closer to 1.3~Gyr and fewer systems that are older.  This may suggest that the underlying mechanism for perturbing planetesimals is related to the {\em total} age of the planetary system, rather than the time from when the mass of the central star changed by a factor of $>$2, consistent with our IMMP model, which depends not only on perturbations, but the total mass of asteroids available as a reservoir.
The relatively large ages at which disks and pollution appear suggests that the mechanism for delivering small planetesimals also must be efficient over Gyr timescales.  Our simulations show that the IMMP model should be effective at delivering asteroids for hundreds of Myr.  

\begin{figure}
\plotone{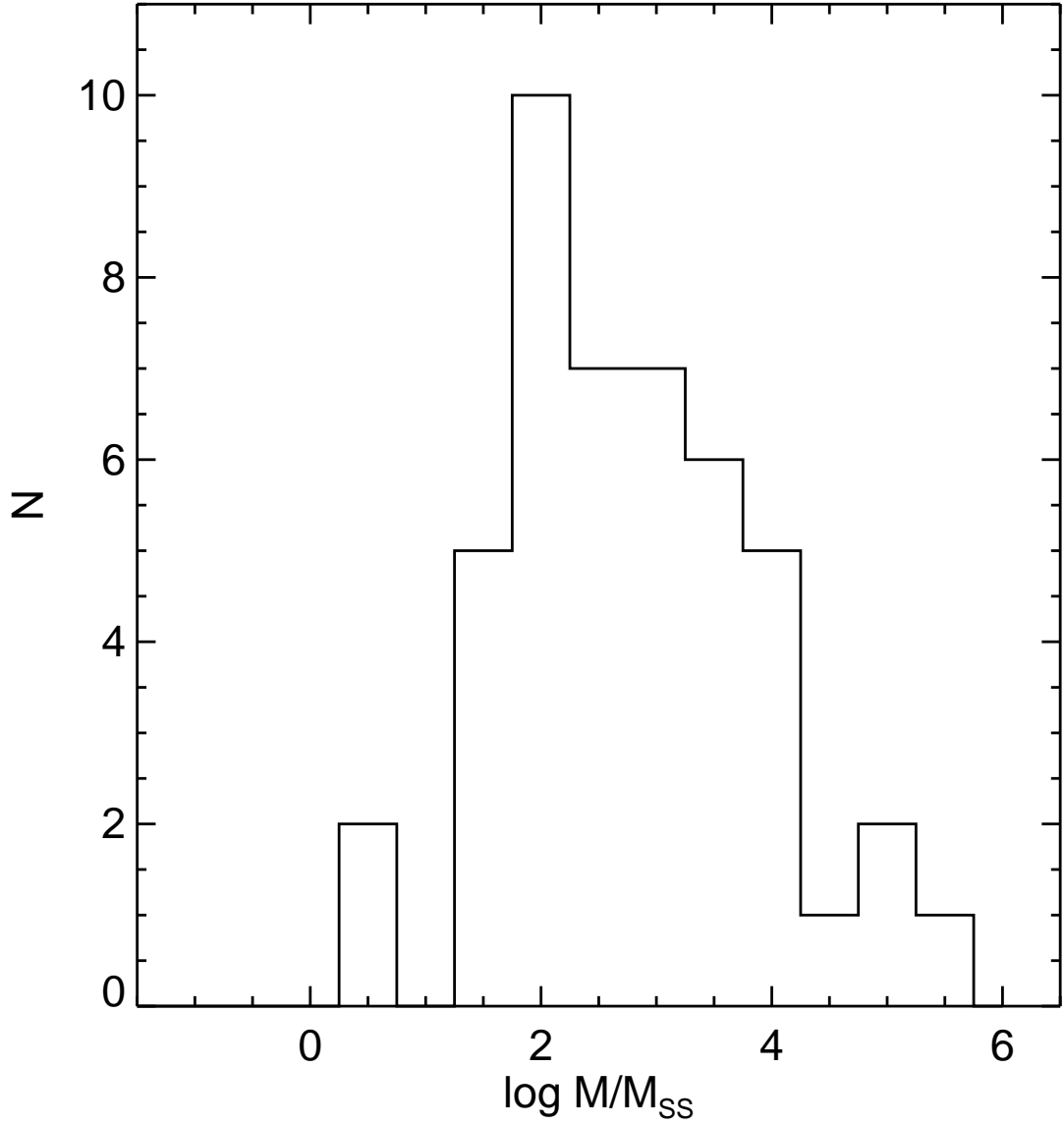}
\caption{\label{fig:f6} Distribution of inferred asteroid belt masses in units of the mass of the Solar System's asteroid belt, M$_{SS}$.  Such a distribution could be used to estimate the amount of dust present around main sequence stars targeted for terrestrial planet searches.  The median value of all masses is 820~M$_{SS}$.}
\end{figure}

Finally, we can use our IMMP results to infer a distribution of asteroid belt masses based on the observed accretion rates for WDs.  We assumed that each non-dusty WD system represents an "average" accretion rate that can be directly compared to the fit of our simulated mass accretion rates.  We then determined a scaling factor between our calculated accretion rate and that observed to infer a total belt mass using Equation \ref{eq:assump}.  The histogram of resulting values is seen in Figure \ref{fig:f6}.  This may overestimate the masses in a particular system if the time between tidal disruptions is longer than the gaseous lifetime of a disk, as a disk evolves from higher accretion rates to lower ones.  The range in asteroid belt masses goes from  4~M$_{SS}$ to 6$\times$10$^{5}$~M$_{SS}$.  The median mass is 820~M$_{SS}$.  This higher mass is may not be unusual given that the median progenitor mass of the non-dusty WDs is 2\Msun, and the median total age of these systems is 1.5~Gyr younger than our own Solar System.  The relative youth and greater progenitor mass could correspond to higher belt masses, as could differing dynamical histories.  This distribution is also biased at higher WD T$_{eff}$ to the largest accretion rates, due to limited sensitivity in the optical to Ca and Mg absorption features.  More unbiased studies may find a lower median belt mass.  We can also compare this amount of mass to that in planetesimals assumed for other models used to explain the observed WD accretion rates.  \citet{jura08} assumed an asteroid belt mass of 10$^{25}$~g, and in \citet{bonsor11}, the median mass of their planetesimal belts exterior to a giant planet was 10 $M_\oplus$ or on the order of 1600 times our assumed belt mass.  If our results hold and the IMMP model is most efficient at explaining the observations, then it would suggest that more massive stars have significantly more massive asteroid belts, or that our asteroid belt is depleted relative to the typical planetary system.

\section{Discussion}
\label{sec:disc}
Our simulations suggest a novel and robust way for explaining the presence of metal polluted white dwarfs in addition to the mechanism first invoked by \citet{debes02} and complementary to the scenario suggested by \citet{bonsor11}.  While the Debes \& Sigurdsson and the Bonsor \& Wyatt mechanisms require multiple planets and a relic icy planetesimal disk to reside within the system, the scenario we present requires only one planet roughly the mass of Jupiter as well as a relic asteroid belt similar to the Solar System's.  Furthermore, since we have looked at only larger asteroids (our sample were incomplete for asteroids with R$<$20~km), there should exist a population of (20/$R_{min}$)$^p$ (where $p$ is the power law value for the asteroidal size distribution and R$_{min}$ is the smallest asteroid size that survives post-main sequence evolution) more objects that could be participating in tidal disruptions.  The recent WISE Mission \citep{wright10} for example will be able to better constrain the Solar System's population of asteroids, as well as find new dusty white dwarfs \citep{debes11}.  We have also shown through tidal disruption simulations that significant disruption of smaller asteroids proceeds at radii comparable to where dusty white dwarfs are observed, confirming what had been proposed by \citet{jura03} for G~29-38.
Despite the seeming complexity of such a situation, these scenarios are the best explanations for the observations at this time.  It is then important to identify which mechanism, if any, dominates and whether there are suitable observational tests that could also potentially discriminate between planetary instability, the IMMP, or exterior mean motion resonances.

There are several important implications to our IMMP simulations which will warrant further and deeper study.  These simulations are useful for understanding the frequency of Solar System like asteroid belts as well as their mass, they provide limits to the mass and location of planets around dusty/polluted white dwarfs, and they help to explain the relative frequency of dusty vs. polluted white dwarfs.  

Firstly, if this mechanism is widespread and more common than perturbations due to exterior resonances, this directly tests the terrestrial planetesimal population of main sequence stars with planetary systems.  This is akin to both a determination of $\eta_{\rm planetesimal}$ as well as a test of how well populated these regions are.  Current observations of main sequence stars with warm dust in the terrestrial planet forming region are rare and represent either recent collisions or high mass asteroidal belts \citep[e.g.][and references therein]{chen06,lisse09,currie11}.  Limits on the masses of these types of planetesimal belts will be useful for constraining how dusty main sequence stars are in the terrestrial planet forming regions, a crucial measurement to determine how successful terrestrial planet imaging missions in the future might be \citep{guyon06}.

\acknowledgements
We wish to thank the anonymous referee for greatly enhancing the clarity and quality of this paper.  The research and computing needed to generate astorb.dat were conducted by Dr. Edward Bowell and funded principally by NASA grant NAG5-4741, and in part by the Lowell Observatory endowment.

\bibliography{wd_chap}
\bibliographystyle{apj}

\clearpage

\clearpage

\end{document}